\documentclass[12pt]{article}
\usepackage{epsfig}

\newcommand{\beq}{\begin{equation}}
\newcommand{\eeq}{\end{equation}}
\newcommand{\beqa}{\begin{eqnarray}}
\newcommand{\eeqa}{\end{eqnarray}}
\newcommand{\beqar}{\begin{eqnarray*}}
\newcommand{\eeqar}{\end{eqnarray*}}

\newcommand{\eg}{{\it e.g.,}\ }
\newcommand{\ie}{{\it i.e.,}\ }
\newcommand{\labell}[1]{\label{#1}} 
\newcommand{\reef}[1]{(\ref{#1})}
\newcommand\prt{\partial}

\newcommand\ttau{{\widetilde \tau}}

\newcommand\tD{{\widetilde D}}

\newcommand\tA{{\widetilde A}}

\newcommand\Tr{{\rm Tr}}
\newcommand\STr{{\rm STr}}

\begin{document}

 \vspace*{1cm}

\begin{center}
{\bf \Large
On the effective action of \\
D-brane-anti-D-brane system \\

 }
\vspace*{1cm}

{Mohammad R. Garousi}\\
\vspace*{0.2cm}
{ Department of Physics, Ferdowsi university, P.O. Box 1436, Mashhad, Iran}\\
\vspace*{0.1cm}
{ Institute for Studies in Theoretical Physics and Mathematics
IPM} \\
{P.O. Box 19395-5531, Tehran, Iran}\\
\vspace*{0.4cm}

\vspace{2cm}
ABSTRACT
\end{center}
We examine the proposal for constructing  the effective action of a
$D_p$-brane-anti-$D_p$-brane system from  the
non-abelian extension of tachyon DBI action. We consider two  prescriptions for the trace in the non-abelian tachyon DBI action. The usual trace and the symmetric trace prescription. The former gives an action for the $D_p\bar{D}_p$ system which reduces to the action proposed by A.Sen for coincident branes. The latter gives a different action which is consistent with the S-matrix element calculations. 
\vfill \setcounter{page}{0} \setcounter{footnote}{0}
\newpage

\section{Introduction} \label{intro}
Study of unstable objects in string theory  might shed new light
in understanding properties of string theory in time-dependent
backgrounds \cite{Gutperle:2002ai,Sen:2002in,Sen:2002an,Sen:2002vv,Lambert:2003zr,Sen:2004nf}. Generally
speaking,  source of instability in these processes  is appearance
of some tachyonic  modes  in the spectrum of these 
objects. It  then makes sense to study them in a field
theory which includes those modes. In this regard, it has been
shown by A. Sen that an effective action of  Born-Infeld type
proposed in \cite{Sen:1999md,Garousi:2000tr,Bergshoeff:2000dq,Kluson:2000iy} can capture many properties
of the decay of non-BPS D$_p$-branes in string theory
\cite{Sen:2002in,Sen:2002an}. 

Recently, unstable objects have been used to study spontaneous chiral symmetry breaking  in  holographic model of QCD \cite{Casero:2007ae,Bergman:2007pm,Dhar:2007bz}. In these studies, flavor branes introduced by placing a set of parallel branes and antibranes on a background dual to a confining color theory \cite{Sakai:2004cn}. 
Detailed study of brane-antibrane system
 reveals   when brane separation is smaller than the
string length scale,  spectrum of this system has two tachyonic
modes \cite{Sen:1998ii}.  The
 effective action  should then include
these tachyonic modes because they are the most important modes
which rule  the dynamics of   the system.  

A proposal by A. Sen for $D_p\bar{D}_p$
effective action when branes are coincident is \cite{Sen:2003tm} 
\beqa
S&=&-\int d^{p+1}\sigma V(|\tau|)\left(\sqrt{-\det\textbf{A}^{(1)}}+\sqrt{-\det
\textbf{A}^{(2)}}\right)\,,\labell{action21}\eeqa where \beqa
\textbf{A}_{\mu\nu}^{(n)}&=&\eta_{\mu\nu}
+2\pi\alpha'F^{(n)}_{\mu\nu}+
\pi\alpha'\left(D_{\mu}\tau(D_{\nu}\tau)^*+D_{\nu}\tau(D_{\mu}\tau)^*\right)\,.\eeqa
which is a generalization of tachyon DBI action \cite{Sen:1999md,Garousi:2000tr,Bergshoeff:2000dq,Kluson:2000iy}. This action has a vortex solution whose world-volume action is given by the action of  stable $D_{p-2}$-brane \cite{Sen:2003tm}. In order to extend the above action to the action of non-coincident branes, it has been proposed   in \cite{Garousi:2004rd} that the  effective action of  $D_p\bar{D}_p$ might  be derived from the effective action of two non-BPS D-branes by projecting it with $(-1)^{F_L}$ where $F_L$ is the spacetime left-handed fermion number. Two non-BPS branes, on the other hand,  may be described effectively by the non-abelian generalization of the tachyon DBI action. This action should  extend the abelian $U(1)$ gauge symmetry  of one non-BPS brane to non-abelian $U(2)$ symmetry of two non-BPS D-branes\cite{Garousi:2000tr}. The non-abelian action can then be found by  converting the  open string fields  to matrix form, changing the ordinary derivative to covariant derivative and performing a trace over the matrices. Various trace prescriptions give different non-abelian theories. In this paper, we would like to consider  two  trace prescriptions. Ordinary trace and symmetric  trace prescription.  In the first case, we shall show that the resulting action  is consistent with the above action when branes are coincident. In the second case, we  shall show that the action is not consistent with the above action. However, the good point about this latter action is that it is consistent with the S-matrix element calculations.

In the next section we shall find the two effective actions for  the
$D_9\bar{D}_9$ system by projecting   the Chan-Paton factors of
the  open string fields in the non-abelian tachyon DBI
action of two non-BPS branes with $(-1)^{F_L}$.  In this section, we compare the two proposal for the effective action and show that they are not the same action. In particular, the symmetric trace action has coupling between $F^{(1)}$ and $F^{(2)}$ whereas there is no such couplings in the ordinary trace action.  In  section 3, we will find the effective actions of the
$D_p\bar{D}_p$ system  for $p<9$ by using the consistency of the effective actions with T-duality transformations. In this section we will show that even the  $D_p\bar{D}_p$ tachyon potential is different in the two effective actions.

\section{$D_9\bar{D}_9$ effective action}

The   effective action for describing the 
dynamics of one non-BPS D$_p$-brane, and its coupling to gravity
and  world-volume gauge field is given by \cite{Sen:1999md,Garousi:2000tr,Bergshoeff:2000dq,Kluson:2000iy}:
 \beqa
S&=&-\int d^{p+1}\sigma V(T)
e^{-\Phi}\sqrt{-\det(P[g_{ab}+B_{ab}]+2\pi\alpha'F_{ab}+2
\pi\alpha'\prt_a T\prt_b T)} \,\,,\labell{dbiac2}\eeqa where
$V(T)$ is the tachyon potential. 
Here $g_{ab} , B_{ab},\Phi$ and $A_a$  are the spacetime metric,
antisymmetric Kalb-Ramond tensor, dilaton and the gauge field,
respectively. In above action $P[\cdots]$ is also the pull-back
of the closed string fields. For
 example,  $P[\eta_{ab}]=\eta_{\mu\nu}\prt_a X^{\mu}\prt_b X^{\nu}=
 \eta_{ab}+\prt_a X^i\prt_b X_i$ in the
static gauge\footnote{Our index convention is that
$\mu,\nu,...=0,1,...,9$; $a,b,...=0,1,...,p$  and
$i,j,...=p+1,...,9$.}. There are different proposal for the tachyon potential \cite{Buchel:2002tj,Lambert:2003zr}. The tachyon potential which is consistent with S-matrix element calculation is $V(T)=T_p(1+\pi\alpha' m^2T^2+\frac{1}{2!}(\pi\alpha' m^2T^2)^2+O(T^6))$ where $m^2=-1/(2\alpha')$. This potential is also consistent with the potential in boundary superstring field theory \cite{Kutasov:2000aq}.

Now consider $N=2$ non-BPS $D_p$-branes. They  may be described
effectively by  non-abelian extension of the above action. To find the non-abelian action for $p<9$, one may consider first the non-abelian action for $p=9$ case which has no transverse scalar field,  and then use the T-duality transformations to find the effective action for any $p$. We consider the following two non-abelian extensions:
\beqa
S_1&\!\!\!=\!\!\!&-\Tr\int d^{10}\sigma V(T)e^{-\Phi}\sqrt{-\det(g_{\mu\nu}+B_{\mu\nu}+2\pi\alpha'F_{\mu\nu}+
\pi\alpha'[D_{\mu} T D_{\nu}T+D_{\nu}TD_{\mu}T])} \labell{dbiac21}\eeqa where we have written the kinetic term in the symmetric form to make the integrand a Hermitian matrix\footnote{Another nonabelian extension of action \reef{dbiac2} has been considered in \cite{Garousi:2004rd} in which the trace has been taken to be the ordinary trace and the kinetic term has been written in symmetric form at the end  after applying the T-duality transformation.}, and
\beqa
S_2&\!\!\!=\!\!\!&-\STr\int d^{10}\sigma V(T)e^{-\Phi}\sqrt{-\det(g_{\mu\nu}+B_{\mu\nu}+2\pi\alpha'F_{\mu\nu}+
2\pi\alpha'D_{\mu} T D_{\nu}T}) \labell{dbiac22}\eeqa 
here the symmetric trace make the integrand to be a Hermitian matrix. In above,  the gauge field
strength and covariant derivative of the
tachyon are\beqa
F_{\mu\nu}&=&\prt_{\mu}A_{\nu}-\prt_{\nu}A_{\mu}-i[A_{\mu},A_{\nu}]\,,\nonumber\\
D_{\mu}T&=&\prt_{\mu}T-i[A_{\mu},T]\,.\nonumber\eeqa Obviously both  actions
\reef{dbiac21} and \reef{dbiac22} have $U(2)$ gauge symmetry and reduce to \reef{dbiac2} for $N=1$. The trace in $S_1$ is the usual trace whereas the trace in $S_2$ is the symmetric trace. That is,  one has to first  expand the square root and the tachyon potential and then make each term of the expansion completely symmetric between all non-abelian expressions
of the form $F_{\mu\nu}, \,D_{\mu}T$ and  the individual
$T$ of the tachyon potential. Only after this rearrangement, one has to perform the trace.   Various couplings in the action \reef{dbiac22} are consistent with the  appropriate disk level S-matrix elements in string
theory \cite{Garousi:2000tr,Garousi:2002wq,BitaghsirFadafan:2006cj}. In particular, the calculation in \cite{BitaghsirFadafan:2006cj} shows that the consistency is hold only if one uses the symmetric trace prescription. 

The proposal for the effective action of  $D_9$-brane anti-$D_9$-brane system \cite{Garousi:2004rd} is to   project   the effective action of two non-BPS $D_9$-brane with $(-1)^{F_L}$.     All fields in the non-abelian tachyon DBI action are invariant under the $(-1)^{F_L}$ projection. However, the Chan-Paton matrices is not invariant under this projection \cite{Sen:1999mg}. It projects the Chan-Paton matrices of two non-BPS $D_9$-brane to the  following matrices: \beqa A_{\mu}=\pmatrix{A_{\mu}^{(1)}&0\cr
0&A_{\mu}^{(2)}},\,\,T=\pmatrix{0&\tau\cr \tau^*&0}\,. \labell{M11}\eeqa
 The superscripts $(1)$ and $ (2)$
refer to the open string fields with both ends on brane $1$ and
$2$, respectively. $\tau (\tau^*)$ refers to the tachyon with one
end on brane $1(2)$ and the other end on brane $2(1)$. Since
there is no off-diagonal terms for the gauge field, the theory
has gauge symmetry $U(1)\times U(1)$.  For  above matrices, one finds
\beqa
F_{\mu\nu}=\pmatrix{F^{(1)}_{\mu\nu}&0\cr 
0&F^{(2)}_{\mu\nu}},\,\,
D_{\mu}T=\pmatrix{0&D_{\mu}\tau\cr 
(D_{\mu}\tau)^*&0}\, \labell{M12} \eeqa 
where $F^{(i)}_{\mu\nu}=\prt_{\mu}A^{(i)}_{\nu}-\prt_{\nu}A^{(i)}_{\mu}$ and $D_{\mu}\tau=\prt_{\mu}\tau-i(A^{(1)}-A^{(2)})\tau$. 

Now one has to perform the traces. Since the matrices $F_{\mu},\,D_{\mu}T$ and $T$ do not commute, one can not perform the trace in $S_2$ without expanding the square root and the tachyon potential. The trace in $S_1$, on the other hand,   is ordinary trace and can be performed for the above matrices without expanding the action. Now using the fact that the tachyon potential is an even function of $T$, one finds that for the above matrices the tachyon potential becomes $V(T)=V(|\tau|)I$ and the covariant derivative term in the action \reef{dbiac21} becomes $ D_{\mu} T D_{\nu}T+D_{\nu}TD_{\mu}T=(D_{\mu} \tau (D_{\nu}\tau)^*+(D_{\mu}\tau)^* D_{\nu}\tau)I$ . Hence, the action \reef{dbiac21} reduces to a diagonal matrix which after performing the trace it becomes
\beqa
S_1&=&-\int d^{10}\sigma V(|\tau|)e^{-\Phi}\left(\sqrt{-\det\textbf{A}^{(1)}}+\sqrt{-\det
\textbf{A}^{(2)}}\right)\,,\labell{action21}\eeqa where \beqa
\textbf{A}_{\mu\nu}^{(n)}&=&g_{\mu\nu}+B_{\mu\nu}
+2\pi\alpha'F^{(n)}_{\mu\nu}+
\pi\alpha'\left(D_{\mu}\tau(D_{\nu}\tau)^*+D_{\nu}\tau(D_{\mu}\tau)^*\right)\,.\eeqa
This action   is the one proposed in \cite{Sen:2003tm}.

Now let us compare the above action with the symmetric trace action \reef{dbiac22} for trivial closed string background \ie $g=\eta,\,B=0,\Phi=0$. Using  the following expansion, one can expand the square root in \reef{action21} and \reef{dbiac22}
 to  produce various interacting
terms \beqa \sqrt{ -\det(M_0+M)}&=&\sqrt{
-\det(M_0)}\left(1+\frac{1}{2}\Tr\left(M_{0}^{-1}M\right)-\frac{1}{4}
\Tr\left(M_{0}^{-1}MM_{0}^{-1}M\right)
\right.\nonumber\\
&+&\left.\frac{1}{8}\left(\Tr\left(M_{0}^{-1}M\right)\right)^2+\frac{1}
{6}\Tr\left(M_{0}^{-1}MM_{0}^{-1}MM_{0}^{-1}M\right)
\right.\nonumber\\
&-&\left.\frac{1}{8}\left(\Tr\left(M_{0}^{-1}M\right)\right)\Tr\left(M_{0}^{-1}MM_{0}^{-1}M\right)
+\frac{1}{48}\left(\Tr\left(M_{0}^{-1}M\right)\right)^3+\cdots\,\right)\nonumber
\label{q13}\eeqa
The terms involving two  gauge fields and two
tachyons are the following: 
\beqa {\cal
L}_1&\!\!\!=\!\!\!&-T_9(2\pi\alpha')\left(m^2|\tau|^2+D\tau\cdot(D\tau)^{*}-\frac{\pi\alpha'}{2}
F^{(1)}\cdot{F^{(1)}} \right)+T_9(\pi\alpha')^3\times\labell{exp1}\\
&&\left(\frac{}{}D\tau\cdot(D\tau)^{*}F^{(1)}\cdot{F^{(1)}}
+m^2|\tau|^2F^{(1)}\cdot{F^{(1)}}-2{F^{(1)}}^{\mu\alpha}F^{(1)}_{\alpha\beta}[D^{\beta}\tau(D_{\mu}\tau)^{*}
+(D^{\beta}\tau)^{*}D_{\mu}\tau ]\right)\nonumber
\eeqa
There are similar terms for $F^{(2)}$.  Note that there is no coupling between $F^{(1)}$ and $F^{(2)}$. 
  The  two  gauge fields and two
tachyons from expanding the action \reef{dbiac22} are
\beqa
{\cal L}_2&=&-T_9(\pi\alpha')\STr\left(m^2T^2+D_{\mu}TD^{\mu}T-\pi\alpha'F_{\mu\nu}F^{\nu\mu}\right)+T_9(\pi\alpha')^3\times\labell{exp2}\\ 
&&\times\STr\left(\frac{}{}D^{\alpha}TD_{\alpha}TF_{\mu\nu}{F}^{\nu\mu}
+m^2T^2F_{\mu\nu}{F}^{\nu\mu}-4{F}^{\mu\alpha}F_{\alpha\beta}D^{\beta}TD_{\mu}T
\right)\nonumber
\eeqa
Writing the symmetric trace in term of ordinary trace, one finds
\beqa
{\cal L}_2&\!\!\!\!=\!\!\!\!&-T_9(\pi\alpha')\Tr\left(m^2T^2+D_{\mu}TD^{\mu}T-\pi\alpha'F_{\mu\nu}F^{\nu\mu}\right)+T_9(\pi\alpha')^3\times\labell{exp2}\\ 
&&\Tr\left(\frac{2}{3}D^{\alpha}TD_{\alpha}TF_{\mu\nu}{F}^{\nu\mu}+\frac{1}{3}D^{\alpha} T F_{\mu\nu}D_{\alpha}TF^{\mu\nu}+\frac{2m^2}{3}T^2F_{\mu\nu}{F}^{\nu\mu}+\frac{m^2}{3} T F_{\mu\nu}TF^{\mu\nu}\right.\nonumber\\
&&\left.\qquad-\frac{4}{3}{F}^{\mu\alpha}F_{\alpha\beta}D^{\beta}TD_{\mu}T-\frac{4}{3}F_{\alpha\beta}{F}^{\mu\alpha}D^{\beta}TD_{\mu}T
-\frac{4}{6}{F}^{\mu\alpha}D^{\beta}TF_{\alpha\beta}D_{\mu}T\right.\nonumber\\
&&\left.\qquad\qquad\qquad\qquad\qquad\qquad\qquad\qquad\qquad\qquad\qquad\qquad-\frac{4}{6}F_{\alpha\beta}D^{\beta}T{F}^{\mu\alpha}D_{\mu}T
\right)\nonumber
\eeqa
Note that the above matrix is Hermitian. Inserting the matrices $T,\,F_{\mu\nu}$ and $D_{\mu}T$ from \reef{M11} and \reef{M12} and performing the trace, one finds
\beqa {\cal
L}_2&\!\!\!=\!\!\!&-T_9(2\pi\alpha')\left(m^2|\tau|^2+D\tau\cdot(D\tau)^{*}-\frac{\pi\alpha'}{2}
\left(F^{(1)}\cdot{F^{(1)}}+
F^{(2)}\cdot{F^{(2)}}\right)\right)+T_9(\pi\alpha')^3\times\nonumber\\
&&\times\left(\frac{2}{3}D\tau\cdot(D\tau)^{*}\left(F^{(1)}\cdot{F^{(1)}}+F^{(1)}\cdot{F^{(2)}}+F^{(2)}\cdot{F^{(2)}}\right)\right.\nonumber\\
&&\left.+\frac{2m^2}{3}|\tau|^2\left(F^{(1)}\cdot{F^{(1)}}+F^{(1)}\cdot{F^{(2)}}+F^{(2)}\cdot{F^{(2)}}\right)\right.\nonumber\\
&&-\left.\frac{4}{3}\left((D^{\mu}\tau)^*D_{\beta}\tau+D^{\mu}\tau(D_{\beta}\tau)^*\right)\left({F^{(1)}}^{\mu\alpha}F^{(1)}_{\alpha\beta}+{F^{(1)}}^{\mu\alpha}F^{(2)}_{\alpha\beta}+{F^{(2)}}^{\mu\alpha}F^{(2)}_{\alpha\beta}\right)\right)
\nonumber
\eeqa
The first line is like the corresponding terms in \reef{exp1}, however, the other couplings are not the same as in \reef{exp1}. In particular, there is coupling between $F^{(1)}$ and $F^{(2)}$. Obviously when tachyon is zero, the two action become identical because the matrix $F$ is diagonal, hence, the symmetric trace and ordinary trace are the same.

\section{$D_p\bar{D}_p$ effective action}

The action for $D_p\bar{D}_p$ system can be found from  $D_9\bar{D}_9$ by using the consistency of the action with T-duality transformations. T-duality transformations in $i=p+1,\cdots, 9$ directions of the $D_9\bar{D}_9$ world volume converts the $D_9\bar{D}_9$ to $D_p\bar{D}_p$, the gauge fields in those directions to $\tA^{(1)}_i=X^{(1)i}/2\pi\alpha'$, $\tA^{(2)}_i=X^{(2)i}/2\pi\alpha'$ and leave unchanged the tachyons. The T-duality of $S_2$ is the non-abelian tachyon DBI action that has been found in \cite{Garousi:2000tr}. The corresponding action for $D_p\bar{D}_p$ is
 \beqa
S_2&=&-\int
d^{p+1}\sigma \STr\left(V(T)\sqrt{\det(Q^i{}_j)}\right.\labell{nonab}\\
&&\times\left.
e^{-\Phi}\sqrt{-\det(P[E_{ab}+E_{ai}(Q^{-1}-\delta)^{ij}E_{jb}]
+2\pi\alpha'F_{ab}+T_{ab})} \right)\,\,,\nonumber \eeqa where
$E_{\mu\nu}=g_{\mu\nu}+B_{\mu\nu}$. The  indexes in this action
are raised  and lowered by $E^{ij}$ and $E_{ij}$, respectively.
The matrices $Q^i{}_j$ and $T_{ab}$ are  \beqa
Q^i{}_j&=&I\delta^i{}_j
-\frac{1}{2\pi\alpha'}L^iL^kE_{kj}\,\,,\labell{mq}\\
T_{ab}&=&2\pi\alpha'D_aTD_bT+D_aTL^i(Q^{-1})_{ij}L^j
D_bT\nonumber\\
&&+iE_{ai}(Q^{-1})^i{}_jL^jD_bT-iD_aTL^i(Q^{-1})_i{}^jE_{jb}
\nonumber\\
&&
+i\prt_aX^i(Q^{-1})_{ij}L^jD_bT-iD_aTL^i(Q^{-1})_{ij}
\prt_bX^j\,\,. \nonumber\eeqa The trace in the action \reef{nonab}
should be completely symmetric between all  matrices
of the form $F_{ab},\prt_aX^i,D_aT, L^i$, individual
$T$ of the tachyon potential and individual $X^i$ of the Taylor
expansion of the closed string fields in the action\cite{Garousi:1998fg}. The matrices $F_{ab},\,D_aT$ and $T$ are those  appear in \reef{M11} and \reef{M12}, and the matrices $\prt_aX^i$ and $L^i$  are
\beqa
\prt_a X^i=\pmatrix{\prt_a X^{(1)}&0\cr 
0&\prt_a X^{(2)}},\,\,
L^i=[X^i,T]=\ell^i\pmatrix{0&\tau\cr 
-\tau^*&0}\, \labell{M13} \eeqa 
where $\ell^i=X^{(1)i}-X^{(2)i}$ is the distance between the two branes.

The T-duality of $S_1$ is obtained by  performing the T-duality for each term of \reef{action21} which has no matrix.
Under T-duality transformation, the covariant derivative of tachyon becomes $\tD_i\ttau=-i\ell^i\tau/2\pi\alpha'$. Using the same steps as those in \cite{Myers:1999ps,Garousi:2000tr} for finding the T-dual action \reef{nonab}, one finds
 \beqa
S_1&\!\!\!=\!\!\!&-\int d^{p+1}\sigma
V(|\tau|)\sqrt{\det(Q)}e^{-\Phi}\left(\sqrt{-\det
\textbf{A}^{(1)}}+\sqrt{-\det
\textbf{A}^{(2)}}\right),\labell{action1}\eeqa where \beqa
\textbf{A}^{(n)}_{ab}&=&P^{(n)}[E_{ab}-\frac{|\tau|^2}
{2\pi\alpha'\det(Q)}E_{ai}\ell^i\ell^jE_{jb}]+
2\pi\alpha'F_{ab}^{(n)}\nonumber\\
&&+\frac{1}{\det(Q)}\left(\frac{}{}\pi\alpha'
\left[D_a\tau (D_b\tau)^*+D_b\tau(D_a\tau)^*\right]\right.\nonumber\\
&&+\frac{\ell\cdot\ell}{4}[\tau (D_a\tau)^*+\tau^*D_a\tau][\tau (D_b\tau)^*+\tau^*D_b\tau]\nonumber\\
&&\left.+\frac{i}{2}[E_{ai}+\prt_a
X^{(n)j}E_{ji}]\ell^i\left[\tau(D_b\tau)^*-\tau^*D_b\tau\right]\right.\labell{A1}\\
&&\left.-\frac{i}{2}\left[\tau (D_a\tau)^*-\tau^* D_a\tau\right]
\ell^i[E_{ib}+E_{ij}\prt_bX^{(n)j}]\right)\,,\nonumber\eeqa
where $\det(Q)=1+|\tau|^2\ell\cdot\ell/2\pi\alpha'$. In the above equation   $P^{(n)}[...]$ means
pull-back of closed string fields on the $n$-th brane, \eg
$P^{(1)}[\eta_{ab}]=\eta_{ab}+\prt_aX^{(1)}_i\prt_bX^{(1)}_j\eta^{ij}$. For simplicity we have assumed that the closed string fields have no $X^i$ dependency\footnote{ The difference between the above action and the one considered in \cite{Garousi:2004rd} is  that the  terms in the third line of \reef{A1} is missing in \cite{Garousi:2004rd}. This results from the different construction of the non-abelian actions. In \cite{Garousi:2004rd}, one first  simplifies $2\pi\alpha'D_aTD_bT+D_aTL^i(Q^{-1})_{ij}L^jD_bT=2\pi\alpha'D_aTD_bT/\det(Q)$ and them, in order to have a real action, one makes the terms $D_aTD_bT$ and $D_aTL^i$  each  symmetric. Whereas in the above action one first makes them  symmetric and then simplifies the result.}.

Now let us compare the  $D_p\bar{D}_p$ potential in the above action with the potential in \reef{nonab}.
  For
small $|\tau|$ the $D_p\bar{D}_p$ potential in the above action has the following expansion: \beqa
&2V(|\tau|)\sqrt{\det(Q)}=&2T_p\left(1+\frac{2\pi\alpha'}{2}\left(\frac{\ell\cdot\ell
 }{(2\pi\alpha')^2}-
\frac{1}{2\alpha'}\right)|\tau|^2-\frac{1}{8\alpha'}\ell^2\tau^4+\cdots\right)\labell{v1}\,.\eeqa The second term in the second parentheses
above  is the mass squared of the tachyon and the first term is
the mass squared of the string   stretched between  two branes,
\ie (tension)$^2\times$(length)$^2$. Note that potential had local
minimum at $|\tau|=0$ only when $\ell>\sqrt{2\pi^2\alpha'}$. The corresponding terms in the action $S_2$ is
\beqa
\STr\left(V(T)\sqrt{\det(Q^i{}_j)}\right)&=&2T_p\left(1+\frac{2\pi\alpha'}{2}\left(\frac{\ell\cdot\ell
 }{(2\pi\alpha')^2}-
\frac{1}{2\alpha'}\right)|\tau|^2-\frac{1}{24\alpha'}\ell^2\tau^4+\cdots\right)\labell{v2}\,.\eeqa 
   While the first three terms in \reef{v1} and \reef{v2} are the same, the last terms are  different. This results  from the symmetric trace prescription, \ie $\STr(T^2LL)=\frac{2}{3}\Tr(TTLL)+\frac{1}{3}\Tr(TLTL)$. The sign of the two terms are different after performing the trace, whereas in \reef{v1} both have the same sign. For other terms, the coefficient of $\tau^n$ is the same for both potentials, however, because of the symmetric trace, the coefficient of $\ell^m\tau^n$ in \reef{v2} is smaller than  the corresponding term in \reef{v1}. 

As we mentioned in the Introduction section, a good feature of $S_1$ is that it has a vortex solution whose world volume action is given by the DBI action of stable $D_{p-2}$-brane \cite{Sen:2003tm}. In that calculation the ansatz for the fields is that $F^{(1)}=-F^{(2)}$, $T\neq 0$ and all other fields are zero. For this assumption the two actions $S_1$ and $S_2$ are not identical. So the vortex solution of $S_1$ is not a solution of $S_2$. On the other hand, if $S_2$ is going to be the effective action of brane-antibrane system, it should have vortex solution. It would be interesting to find such solution.



\end{document}